\documentstyle[aps,12pt]{revtex}
\begin{document}
\title{Energy band structure and intrinsic coherent properties in two weakly linked
Bose Einstein Condensates}
\author{Wei-Dong Li$^1$, Yunbo Zhang$^{1,2}$, J. -Q. Liang$^1$}
\address{$^1$ Department of Physics and Institute of Theoretical Physics, Shanxi\\
University, Taiyuan 030006, China\\
$^2$Department of Physics, University of Turku, FIN-20014 Turun yliopisto,\\
Finland}
\date{\today}
\maketitle

\begin{abstract}
The energy band structure and energy splitting due to quantum tunneling in
two weakly linked Bose-Einstein condensates were calculated by using the
instanton method. The intrinsic coherent properties of Bose Josephson
junction were investigated in terms of energy splitting. For $E_C/E_J\ll 1$,
the energy splitting is small and the system is globally phase coherent. In
the opposite limit, $E_C/E_J\gg 1$, the energy splitting is large and the
system becomes a phase dissipation. Our reslults suggest that one should
investigate the coherence phenomna of BJJ in proper condition such as $%
E_C/E_J\sim 1$.

\end{abstract}

\section{Introduction}

Two weakly linked Bose-Einstein condensates (BECs) behave as the
superconductor Josephson junction. This remarkable feature has been
investigated both theoretically \cite
{clarke,smerzi,smerzi1,niu,smerzi3,stringari2} and experimentally\cite
{stringari,stringari1,anderson1,burger,smerzi2}. The existence of a
Josephson current through a potential barrier between two superconductors or
between two superfluids is a direct manifestation of macroscopic quantum
coherence\cite{anderson}. The experimental realization of Bose-Einstein
condensation (BEC) of weakly interacting alkali atoms \cite{stringari} has
provided a route to study neutral superfluids in a controlled and tunable
environment \cite{stringari1}. The possibility of loading a BEC in a
one-dimensional periodic potential has allowed the observation of quantum
phase effects on a macroscopic scale such as quantum interference \cite
{anderson1}, superfluidity on a local scale \cite{burger} and an oscillating
atomic current in Josephson junction arrays \cite{smerzi2}.

The analogy of the voltage-current characteristic in superconductor
Josephson junction was proposed theoretically in Bose-Josephson junction 
\cite{smerzi,smerzi1,niu,smerzi3}. The macroscopic BEC's coherence has been
demonstrated by interference experiments \cite{hall}, and the first evidence
of the coherent tunneling in an atomic array, related to the ``ac''
Josephson effect, has been reported \cite{anderson1}. A ``dc''
current-biased Bose-Josephson junction can be simulated by a tunneling
barrier moving with constant velocity across the trap\cite{smerzi1}. At a
critical velocity of the barrier (proportional to the critical tunneling
current), a sharp transition between the ``dc'' and ``ac'' Josephson regimes
was predicted. Thus two weakly linked condensates exhibit the analog of the
resistively shunted superconductor Josephson junctions. The ``secondary''
quantum macroscopic effects in small capacitance Josephson junction had
attracted much attention both from theorists and experimentalists for decades%
\cite{zaikin,apenko,zorin,averin}. These phenomena manifest the quantum
behavior of a Josephson junction as a macroscopic object, in contrast with
such ``primary'' quantum macroscopic phenomena as the Josephson junction
itself\cite{zaikin}. In these cases, quantum fluctuations of the phase
difference $\varphi $ across the junctions become important. This
necessitates treating the phase as a quantum operator $\hat{\varphi}$, which
is canonically conjugate to the operator $\hat{N}$. Novel macroscopic
quantum phenomena, such as ``Bloch'' oscillaition, had been reported in
current-biased Josephson junctions\cite{zaikin}. To understand these effects
and present a general picture of the low temperature dynamics of Josephson
junctions, a simple theory\cite{apenko,zorin,averin} has been suggested
based on the extended coordinates, $\varphi \in \left[ -\infty ,+\infty
\right] $.

More recently, the first experimental of BJJ have been realized in a purely
magnetic double-well potential\cite{walraven}. It will stimulate the further
study on the BJJ. Then, can we investigate the ``secondary'' quantum
macroscopic effects\cite{zaikin} in two weakly linked BECs? It was argued in 
\cite{smerzi,smerzi1,stringari2} that the dynamics of the phase of this
system can be mapped onto the sine-Gordon quantum mechanical Hamiltonian (%
\ref{sh}). In this paper, we calculated the energy band structure, ``Bolch''
wave and the energy splitting of the Bose-Josephson junction due to the
quantum tunneling using the instanton method\cite{coleman}. As a simply
application, we investigate the coherent properties of the Bose-Josephson
junction. The results from this method agree exactly with the prediction of
Stringari \cite{stringari2}.

\section{Model and its instantons solution}

All quantities describing the junction should be considered as operators
rather than the classical variables. The operators corresponding to the main
variables, the phase difference $\varphi $ and the particle number $N$ of
the Bose-Josephson junction, satisfy the commutation relation: $[\varphi
,N]=i$, so that $\varphi $ and $N$ are related by the Heisenberg uncertainty
relation which was disregarded by the ``classical'' theory of the
Bose-Josephson junction \cite{smerzi}.

The idealization of two weakly linked BECs can be described as a two-mode
bosonic system \cite{smerzi}. In the ``phase'' representation the relevant
quantum observables are the difference of phases and number of atoms between
the two condensates in each trap. The Hamiltonian can be written in terms of
a ``quantum pendulum'' (Mathieu) equation 
\begin{equation}
\widehat{H}=-\frac{E_C}2\frac{\partial ^2}{\partial \phi ^2}+E_J\cos \phi .
\label{sh}
\end{equation}
The ``charging energy'' $E_C$ and the ``Josephson coupling energy'' $E_J$
can be calculated as overlap integrals \cite{smerzi,smerzi1}, 
\begin{eqnarray}
E_C &=&2g\int dr\Phi _1^4(r)=2g\int dr\Phi _2^4(r),  \nonumber \\
E_J &=&N\int dr\Phi _1^{*}(r)[-\frac{\hbar ^2}{2m}\nabla ^2+V_{ext}+\frac{gN}%
2(\Phi _1^2(r)+\Phi _2^2(r))]\Phi _2(r)
\end{eqnarray}
with the one-body wave functions $\Phi _1(r)$, $\Phi _2(r)$ localized in the
trap $1$, $2$; and $\int dr\Phi _1^{*}(r)\Phi _2(r)=0$, $\int dr\Phi
_{1,2}^{*}(r)\Phi _{1,2}(r)=1$, $g=\frac{4\pi \hbar ^2a}m$; $a$ is the
scattering length and $m$ the atomic mass; $N=\widehat{a}_1^{\dagger }%
\widehat{a}_1+\widehat{a}_2^{\dagger }\widehat{a}_2$ is the total number of
atoms.

For convenience, we modify the potential as $U_J(\phi )=E_J(1+\cos \phi )$.
Then, the Hamiltonian $\widehat{H}$ in Eq. (\ref{sh}) can be considered as
one-dimensional quantum particles with mass $m_\phi =\frac{\hbar ^2}{E_C}$
moving along the $\phi $ axis in sine-Gordon potential $U_J(\phi )$. Due to
the translational symmetry ($\phi ->\phi +2\pi $) of the Hamiltonian (\ref
{sh}), the set of the eigenfunctions should include the Bloch wave functions 
\begin{eqnarray}
\psi _n\left( \phi \right) &=&u_{n,q}\left( \phi \right) \exp \left( iq\phi
\right) ,\qquad n=0,1,2,...,  \nonumber \\
u_{n,q}\left( \phi \right) &=&u_{n,q}\left( \phi +2\pi \right) ,\qquad
-\infty <q<\infty  \label{ef}
\end{eqnarray}
where $q$ is an arbitrary (real) constant vector. Substitution of the wave
function (\ref{ef}) in the Schr\"{o}dinger equation leads immediately to the
picture of band energy spectrum and relative well known effects in solid
state theory. From Ref. \cite{zorin,averin}, this is nothing but Mathieu
equation, so that these functions can be readily calculated. Some of their
asymptotic properties (in tight-binding limit and weak-binding limit) can be
expressed analytically.

The energy band structure and energy splitting for the sine-Gordon potential
can be calculated alternatively by instanton method\cite{liang,liang1}. The
advantage of this nonperturbative method, as presented here, is that it
gives not only a more accurate description of the tunneling phenomena but
also a comprehensive physical understanding in the context of quantum field
theory.

The effective Lagrangian is 
\begin{equation}
L=\frac 12m_\phi \left( \frac{d\phi }{dt}\right) ^2-U_J\left( \phi \right) .
\end{equation}
The classical solution which extremizes the action is seen to satisfy the
equation of motion 
\begin{equation}
\frac 12m_\phi \left( \frac{d\phi _c}{d\tau }\right) ^2+U_J\left( \phi
_c\right) =-E_{cl},  \label{ss}
\end{equation}
where the Wick rotation $\tau =it$ has taken the system into Euclidean time.
Equation (\ref{ss}) can be regarded as the equation of motion of a
pseudo-particle with the classical energy $E_{cl}\geq 0$, which is a
constant of integration. With $E_{cl}$ being confined to a region $0\leq
E_{cl}\leq E_J$, the configuration $\phi _c$ becomes periodic such that $%
\phi _c(\tau +{\cal T})=\phi _c(\tau )$, which now corresponds to the
periodic boundary condition in the space coordinate \cite{liang}. The
classical solution is 
\begin{equation}
\phi _c=2\arcsin [k\text{sn}(\omega _0(\tau +\tau _0))],  \label{sl}
\end{equation}
where $\omega _0=\sqrt{E_JE_c}/\hbar $ is the classical plasma frequency 
\cite{stringari2}, sn is the Jacobian elliptic function with modulus $k=%
\sqrt{1-E_{cl}/2E_J}$. The elliptic function sn$(\omega _0(\tau +\tau _0))$
has a period $4{\cal K}(k)$ with ${\cal K}(k)$ the complete elliptic
integral of first kind. For zero energy $E_{cl}=0$ $(k\rightarrow 1)$, the
periodic solution reduces to the vacuum instanton configuration $\phi
_c\rightarrow 2\arcsin [\tanh (\omega _0(\tau +\tau _0))]$.

\section{Energy Band structure and the transition amplitude for quantum
tunneling}

The sine-Gordon potential has an infinite number of degenerate vacua.
Quantum tunneling between neighboring vacua leads to the level splitting,
while the levels extend to bands due to the translational symmetry expressed
by $U_J(\phi +2\pi )=U_J(\phi )$. In the narrow-band approximation one
finds, for the energy, the expression 
\begin{equation}
E=E_i+\sum_nJ(R_m-R_n)e^{iq(R_m-R_n)},
\end{equation}
where $E_i$ denotes the $i$th eigenvalue of the energy in each well for the
harmonic oscillator potential $U(\phi )=\frac 12(\phi -R_n)^2$, $R_n=2n\pi $
is the position of the $n$th minimum, $J(R_{n^{\prime }}-R_n)\!\equiv \!\int
\psi _i^{*}(\phi \!-\!R_{n^{\prime }})[U_J(\phi )\!-\!U(\phi )]\psi _i(\phi
\!-\!R_n)d\phi $ is the overlap integral, $\psi _i$ is the eigenfunction
corresponding to eigenvalue $E_i$, and $q$ is the Floquet parameter
associated with the Bloch wave function. If only the contribution from the
nearest neighbors is taken into account, i.e., $J(R_{n^{\prime }}-R_n)$ for $%
\mid n^{\prime }-n|>1$ is taken to be zero, the energy band formula reduces
to 
\begin{equation}
E=E_i+2J\cos (2\pi q).
\end{equation}

The parameter $J$ is just the level splitting resulting from quantum
tunneling (the wave functions are periodic for $q=0$). We will consider the
case of potential wells surrounded by very high potential barriers with
correspondingly small tunneling contribution to the eigenvalues. They are
almost those of degenerate harmonic oscillators, and in this asymptotic case
we are not concerned with the entire bands but only with their edges which
correspond to alternately even and odd states. Then, we suppose $\mid
i\rangle _R$, $\mid i\rangle _L$ are degenerate eigenstates in neighboring
wells, respectively with the same energy eigenvalue $E_i$ such that $H^0\mid
i\rangle _{R,L}=E_i\mid i\rangle _{R,L}$ where $H^0$ is the Hamiltonian of
the harmonic oscillator as the zero order approximation of the system. The
degeneracy will be removed by the small tunneling effect which leads to the
level splitting. The eigenstates of the Hamiltonian $H$ become 
\begin{equation}
\mid i\rangle _o=\frac 1{\sqrt{2}}\left( \mid i\rangle _R-\mid i\rangle
_L\right) ,\qquad \mid i\rangle _e=\frac 1{\sqrt{2}}\left( \mid i\rangle
_R+\mid i\rangle _L\right)
\end{equation}
with eigenvalues $E_i\pm \Delta E_i$, respectively. $\Delta E_i$ denotes the
shift of one oscillator level. It is obvious that $_R\langle i\mid H-H^0\mid
i\rangle _L=2J=\Delta E_i$. In the following we calculate this energy shift $%
\Delta E_i$ as resulting from periodic instantons and
instanton-anti-instanton pairs.

The amplitude for a transition from one well to its neighboring well at the
energy $E_i$ due to instanton tunneling can be written as 
\begin{equation}
A_{+,-}=_{+}\langle E_i|e^{-2\frac{\widehat{H}T}\hbar }|E_i\rangle
_{-}\simeq \pm e^{-2\frac{E_i}\hbar T}\sinh (2\frac{\Delta E_iT}\hbar ),
\label{amm}
\end{equation}
where we neglect overlap of the wave functions which dominate over either
well. The amplitude (\ref{amm}) can also be calculated with the help of the
path-integral method, $A_{+,-}=\int \psi _{E,+}^{*}(\phi _f)K(\phi _f,\tau
_f;\phi _i,\tau _i)\psi _{E,-}(\phi _i)d\phi _fd\phi _i$, where the Feynman
kernel is defined as usual by 
\begin{equation}
K(\phi _f,\tau _f;\phi _i,\tau _i)=\int_{\phi _i}^{\phi _f}{\cal D}[\phi
]e^{-S/\hbar }  \label{kernel}
\end{equation}
with $\phi _f\equiv \phi (\tau _f)$, $\phi _i\equiv \phi (\tau _i)$ and $%
\tau _f-\tau _i=2T$. What we are interested in is this expression in the
limits $\phi _i\rightarrow -a$, $\phi _f\rightarrow a$ ($\pm a$ are the
turning points), namely, the tunneling propagator through one of the
barriers. $S=\int_{\tau _i}^{\tau _f}[\frac 12m_\phi (\frac{d\phi }{d\tau }%
)^2+U_J(\phi )]d\tau $ is the Euclidean action of the pseudo-particle, and $%
\psi _{E_i,+}(\phi _i)$ ($\psi _{E_i,-}(\phi _i)$) is the wave function of
the right- (left-) hand wells.

The functional integral $K(\phi _f,\tau _f;\phi _i,\tau _i)$ can be
evaluated with the stationary method by expanding $\phi \left( \tau \right) $
about the classical trajectory $\phi _c\left( \tau \right) $ and thus we set 
$\phi \left( \tau \right) =\phi _c\left( \tau \right) +\chi \left( \tau
\right) $, where $\chi \left( \tau \right) $ is the small fluctuation with
boundary conditions $\chi \left( \tau _i\right) =\chi \left( \tau _f\right)
=0$. Substitution of $\phi \left( \tau \right) $ into Eq.(\ref{kernel}) and
keeping only terms containing $\chi \left( \tau \right) $ up to the one-loop
approximation yields $K(\phi _f,\tau _f;\phi _i,\tau _i)=\exp [-\frac{%
S_c(\tau )}\hbar ]I$, where $I=\int_{\chi \left( \tau _i\right) =0}^{\chi
\left( \tau _f\right) =0}$ ${\cal D}[\chi ]e^{-\delta S/\hbar }$ is the
fluctuation function integral with the fluctuation action $\delta
S=\int_{\tau _i}^{\tau _f}\chi M\chi d\tau ,$ where $M=-\frac{m_\phi d^2}{%
d\tau ^2}+V^{\prime \prime }\left( \phi _c\left( \tau \right) \right) $ is
the second variational operator of the action. The classical action $%
S_c(\tau )$ is evaluated along the trajectory $\phi _c\left( \tau \right) $
so that 
\begin{eqnarray}
S_c(\tau ) &=&\int_{\tau _i}^{\tau _f}\left[ \frac{m_\phi }2\left( \frac{%
d\phi _c\left( \tau \right) }{d\tau }\right) ^2+V\left( \phi _c\left( \tau
\right) \right) \right] d\tau   \nonumber \\
&=&2ET+\frac{8E_J}{\omega _0}[{\cal E}(k)-k^{\prime 2}{\cal K}(k)],
\end{eqnarray}
where ${\cal E}(k)$ denotes the complete elliptic integral of the second
kind and $k^{\prime 2}=1-k^2$. Following the standard procedure of the
periodic instanton calculation in Refs. \cite{liang} and \cite{liang1}, the
functional intergal $I$ can be wrtten as $I=\frac 1{\sqrt{2\pi }}\left[
N\left( \tau _i\right) N\left( \tau _f\right) \int_{\tau _i}^{\tau _f}\frac{%
d\tau }{N^2\left( \tau \right) }\right] ^{-1/2}$, where $N\left( \tau
\right) =\frac{d\phi _c\left( \tau \right) }{d\tau }$, is the zero eigenmode
of $M$. To obtain the desired results, the contributions from the infinite
number of the instantons and anti-instanton pairs have to be taken into
account. The total amplitude is found to be 
\begin{equation}
A_{+,-}=e^{-2\frac{E_i}\hbar T}\sinh (\frac{\omega _0T}{2{\cal K}(k^{\prime
})}e^{-\frac W\hbar }),  \label{famm}
\end{equation}
with $W=\frac{8E_J}{\omega _0}[{\cal E}(k)-k^{\prime 2}{\cal K}(k)]$.
Comparing (\ref{famm}) and (\ref{amm}) leads to $\Delta E=\frac{\hbar \omega
_0}{4{\cal K}(k^{\prime })}\exp (-\frac W\hbar )$. Rescaling this formula in
the unit of $E_J$, we arrive at our final result. 
\begin{equation}
\frac{\Delta E}{E_J}\!=\!\sqrt{\frac{E_C}{E_J}}\frac 1{4{\cal K}(k^{\prime })%
}\exp [-8\sqrt{\frac{E_J}{E_C}}({\cal E}(k)\!-\!k^{\prime 2}{\cal K}(k))].
\label{split}
\end{equation}

The formula (\ref{split}) shows that the widths of the energy bands are very
sensitive to the dimensionless parameter of $E_C/E_J$, which is a critical
parameter characterizing the dynamical properties of the system such as the
coherence \cite{stringari2} and the transition \cite{zaikin}. A direct
application of this energy splitting is to investigate the coherence
properties of BJJ. Quantum coherence requires that the relative phase of the
order parameter should be preserved over times of the order $\Gamma ^{-1}$,
which is just the tunneling amplitude\cite{leggett}. From the above
calculation, we know that the energy splitting $\Delta E$ is proportional to 
$\Gamma $. Therefore the energy splitting describes the coherent properties
of the system.

Fig. 1 shows the energy splitting $\Delta E/E_J$ as a function of the ratio $%
E_C/E_J$ in the ground state ($E_{cl}=0$). The figure shows that for values
of $E_J$ smaller than $E_C$ the energy splitting is significantly increased,
indicating the occurrence of a continuous transition to the phase
dissipation (the number squeezed regime) \cite{stringari2,stringari3}. In
the limit $E_C/E_J\ll 1$ the system undergoes small oscillations around the
equilibrium. In this limit the correlation between the neighboring wells is
small. One can regard the systems as a globally coherent object described by
a unique order parameter. On the contrary $E_C/E_J\gg 1$ the behavior of the
system is very different. The quantum fluctuation is enhanced due to the
increasing of the tunneling between the neighboring wells in phase space,
showing that the relative phase between two condensates are distributed in a
random way. At the same time, the fluctuation of the relative number of
atoms in two traps becomes smaller and smaller.

It is interesting to investigate the properties of the systems under the
excited states ($E_{cl}\neq 0$) which may be excited by the thermal
fluctuations or other reasons. We plotted the energy splitting as a function
of the dimensionless parameter $E_{cl}/E_J$ for two different values of $%
E_C/E_J$ in Fig. 2. One clearly observes that even if quantum effects are
small the decoherence due to higher excited states may become important. It
is interesting that the respone of the system to the decoherence perform a
total different behavior. States with small quantum effects are more
sensitive to the decoherence fluctuations than those with relatively large
vaules of $E_C/E_J$. The system with little higher $E_C/E_J$ state should be
a good candidate to investigate the quantum coherence phenomena, in $%
E_C/E_J\sim 1$. Because in this case the system with a good coherence will
preserve its coherence for a relative large range of the energy perturbation.

Energy spectrum and energy splitting due to quantum tunneling in BJJ have
been calculated by means of instanton method. Based on this energy splitting
formula, we also investigated the coherence property of BJJ. Our results
agree exactly with that in Ref. \cite{stringari2}. This analysis makes it
possible to investigate the ``secondary'' quantum phenomena in BJJ (see, for
instance, \cite{averin}) and presents a general picture of low temperature
dynamics of Josephson junctions in BEC.

It is great pleasure to thank A. Smerzi and L. P. Pitaeviskii for many
fruitful discussions. This work is supported by NSF of China (Grants
10175039 and 90203007), Science Funding (Grant 20001003) and Youth Funding
(Grant 20011002) from Shanxi Province of China.

Figure Captions

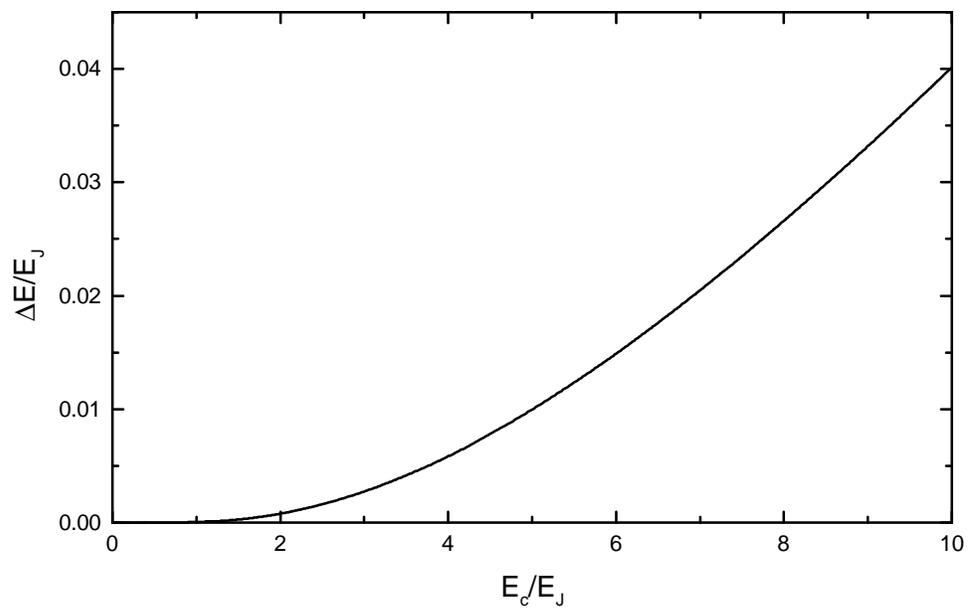
Fig. 1. Ground state energy splitting as a function of the ratio $E_C/E_J$.

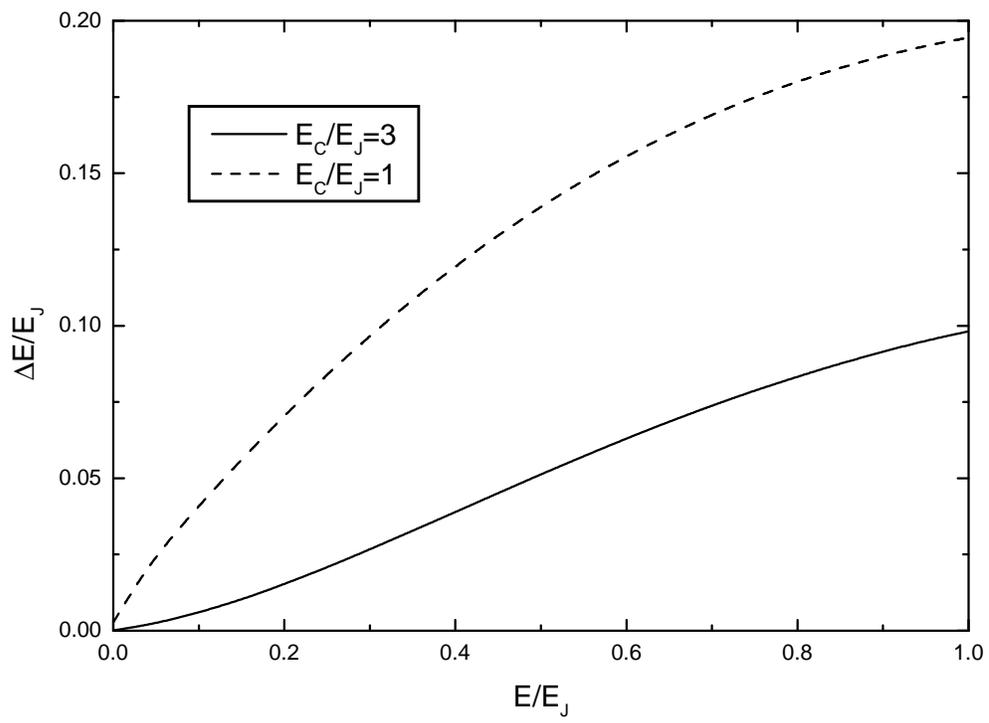
Fig. 2. Energy splitting as a function of the ratio $E/E_J$ for $E_C/E_J=1$
(solid line), $E_C/E_J=3$ (dashed line).

\end{document}